
\input phyzzx

%
\catcode`\@=11 
\def\papersize{\hsize=40pc \vsize=53pc \hoffset=0pc \voffset=1pc
   \advance\hoffset by\HOFFSET \advance\voffset by\VOFFSET
   \pagebottomfiller=0pc
   \skip\footins=\bigskipamount \normalspace }
\catcode`\@=12 
\papers

\def\to{\rightarrow}

\vsize=21.5cm
\hsize=15.cm

\tolerance=500000
\overfullrule=0pt

\pubnum={LPTENS-95/1 \cr
{\tt hep-th@xxx/9501033}\cr
January 1995}

\date={}
\pubtype={}
\titlepage
\title{NON-LOCAL EXTENSIONS OF THE CONFORMAL ALGEBRA :\break
MATRIX $W$-ALGEBRAS, MATRIX KdV-HIERARCHIES \break
AND NON-ABELIAN TODA THEORIES}
\author{{
Adel~Bilal}
}
\address{
{\it CNRS - Laboratoire de Physique Th\'eorique de l'Ecole
Normale Sup\'erieure,}
\foot{
unit\'e propre du CNRS, associ\'e \`a l'Ecole
Normale Sup\'erieure et \`a l'Universit\'e Paris-Sud} \break
{\it 24 rue Lhomond, 75231
Paris Cedex 05, France} \break
{\it e-mail:} {\tt bilal@physique.ens.fr}
}
\vskip 3.mm
\def\d{\partial}
\abstract{
In the present contribution, I report on certain {\it non-linear} and
{\it non-local} extensions of the conformal (Virasoro) algebra. These
so-called $V$-algebras are matrix generalizations of $W$-algebras. First,
in the context of two-dimensional field theory, I discuss the non-abelian
Toda model which possesses three conserved (chiral) ``currents". The
Poisson brackets of these ``currents" give the simplest example of a
$V$-algebra. The classical solutions of this model provide a free-field
realization of the $V$-algebra. Then I show that this $V$-algebra is
identical to the second Gelfand-Dikii symplectic structure on the
manifold of $2\times 2$-matrix Schr\"odinger operators $L=-\d^2+U$ (with
$\tr\sigma_3 U=0$). This provides a relation with matrix KdV-hierarchies
and allows me to obtain an infinite family of conserved charges
(Hamiltonians in involution). Finally, I work out the general
$V_{n,m}$-algebras as symplectic structures based on $n\times n$-matrix
$m^{\rm th}$-order differential operators $L=-\d^m +U_2\d^{m-2}+U_3
\d^{m-3}+\ldots +U_m$. It is the absence of $U_1$, together with the
non-commutativity of matrices that leads to the non-local terms in the
$V_{n,m}$-algebras. I show that the conformal properties are similar to
those of $W_m$-algebras, while the complete $V_{n,m}$-algebras are
much more complicated, as is shown on the explicit example of $V_{n,3}$.
}
\vskip 3.mm
\centerline{\it Invited lecture delivered in Strasbourg, France, December 1994,
 at the }
\centerline{\it ``59. Rencontre entre Physiciens
Th\'eoriciens et Math\'ematiciens",}

\endpage
\pagenumber=1

 \def\PL #1 #2 #3 {Phys.~Lett.~{\bf #1} (#2) #3}
 \def\NP #1 #2 #3 {Nucl.~Phys.~{\bf #1} (#2) #3}
 \def\PR #1 #2 #3 {Phys.~Rev.~{\bf #1} (#2) #3}
 \def\PRL #1 #2 #3 {Phys.~Rev.~Lett.~{\bf #1} (#2) #3}
 \def\CMP #1 #2 #3 {Comm.~Math.~Phys.~{\bf #1} (#2) #3}
 \def\IJMP #1 #2 #3 {Int.~J.~Mod.~Phys.~{\bf #1} (#2) #3}
 \def\JETP #1 #2 #3 {Sov.~Phys.~JETP.~{\bf #1} (#2) #3}
 \def\PRS #1 #2 #3 {Proc.~Roy.~Soc.~{\bf #1} (#2) #3}
 \def\IM #1 #2 #3 {Inv.~Math.~{\bf #1} (#2) #3}
 \def\JFA #1 #2 #3 {J.~Funkt.~Anal.~{\bf #1} (#2) #3}
 \def\LMP #1 #2 #3 {Lett.~Math.~Phys.~{\bf #1} (#2) #3}
 \def\IJMP #1 #2 #3 {Int.~J.~Mod.~Phys.~{\bf #1} (#2) #3}
 \def\FAA #1 #2 #3 {Funct.~Anal.~Appl.~{\bf #1} (#2) #3}
 \def\AP #1 #2 #3 {Ann.~Phys.~{\bf #1} (#2) #3}
 \def\MPL #1 #2 #3 {Mod.~Phys.~Lett.~{\bf #1} (#2) #3}

\def\ap{a_+}
\def\am{a_-}
\def\au{a_1}
\def\ad{a_2}
\def\d{\partial}

\def\du{\partial_u}
\def\dv{\partial_v}
\def\f{\phi}
\def\vf{\varphi}

\def\s{\sigma}
\def\sp{\sigma^+}

\def\l{\lambda}
\def\t{\tau}

\def\is{\int {\rm d}^2 \sigma}

\def\eoms{equations of motion\ }

\def\n{\nu}

\def\e{\epsilon}
\def\es{\epsilon(\s-\s')}
\def\ds{\delta(\s-\s')}
\def\dsp{\delta'(\s-\s')}

\def\dsppp{\delta'''(\s-\s')}
\def\a{\alpha}
\def\b{\beta}
\def\g{\gamma}
\def\gd{\gamma^2}
\def\gmd{\gamma^{-2}}
\def\rd{{\rm d}}
\def\th{{\rm th}}
\def\sh{{\rm sh}}
\def\ch{{\rm ch}}
\def\G{{\cal G}}
\def\la{\vert \l_\a \rangle}
\def\lb{\langle \l_\b \vert}

\def\pr{\Pi_r}
\def\pt{\Pi_t}
\def\pf{\Pi_\f}
\def\Tb{{\bar T}}
\def\Vb{{\bar V}}

{ \chapter{ Introduction}}

In this contribution I will report on certain {\it non-linear} and {\it
non-local} extensions of the conformal or Virasoro algebra
(with central extension). These algebras will be
called $V$-algebras.

 The procedure will be to
go from physics to mathematics. In particular, I will start with a certain
 $1+1$
dimensional field theory (non-abelian Toda field theory)
 defined by its action functional. The
corresponding equations of motion admit three  conserved ``left-moving" and
three conserved ``right-moving" quantities (spin-two currents),
expressed in terms of the
interacting fields. I will define canonical Poisson
brackets from the action functional, which provide the phase space with
the canonical symplectic structure. Then I compute the Poisson brackets
of the conserved quantities. This algebra  closes in a non-linear
and non-local way, reminiscent of $W$-algebras (which are however local),
so it seemed appropriate to call this algebra a $V$-algebra.  The explicit
classical solutions of the equations of motion induce a ``chiral"
realization of the conserved quantities (i.e. of the generators of the
$V$-algebra) in terms of free fields. This will be the contents of section 2.

Then, in the third section, I relate the previously found results to
standard mathematical structures. I show that exactly the same
algebra is obtained as the second Gelfand-Dikii symplectic structure based
on a second-order differential operator, that is a $2\times 2$-matrix
Schr\"odinger operator $L=-\d^2+U$. Applying a straightforward matrix
generalization of the classical work on the resolvent by Gelfand and Dikii,
 this provides us with an infinity of Hamiltonians in involution, hence an
infinity of conserved  charges for the non-abelian Toda field theory I
started with. It also immediately implies the connection with matrix
KdV-hierarchies.

In the fourth section, I generalize all these developments to
$n\times n$-matrix $m^{\rm th}$-order differential operators
$L=-\d^m +U_2\d^{m-2}+U_3\d^{m-3}+\ldots +U_m$ and
so-called $V_{n,m}$-algebras. The latter are $n\times n$-matrix
generalizations of the $W_m$-algebras. They are non-local due to the reduction
to
$U_1=0$ and the
non-commutativity of matrices. This fourth section is mathematically
self-contained,
(except that the reader is referred to my original paper for the proofs). Of
course,
taking this section just by itself, misses  the whole point of the present
contribution (at least in my physicist's point of view), which is to relate
abstract
 mathematical structures to certain physical
theories.

\chapter{The non-abelian Toda field theory}

\section{The action and \eoms}

The one plus one dimensional field theory I will consider is defined by its
action functional
$$ S\equiv S[r,t,\f]= {2\over \gd}\int \rd\s \rd\t
\left( \du r\dv r+\th^2 r\, \du t\dv t +\du \f\dv\f + \ch 2r\, e^{2\f}
\right) \ ,
\eqn\uiii$$
where $\t$ is the time and $\s$ the space-coordinate and
$u=\t+\s$ and $v=\t-\s$.
The  three fields are $r(\t,\s), t(\t,\s)$ and $\f(\t,\s)$. This action
has the following physical interpretation.
The first two, $\f$-independent terms, constitute a sigma-model
describing a string on a two-dimensional black hole background,
 while the last two, $\f$-dependent terms
correspond to an ``internal" field $\f$ (or a flat third dimension) and
a tachyon potential $\ch 2r\, e^{2\f}$. However, this interpretation need
not concern us here. The constant $\gd$ plays the role
of the Planck constant and will be seen later to control the central
charge of the conformal algebra. As discussed below, this model is obtained
by gauging a {\it nilpotent} subalgebra of the Lie algebra $B_2$.
The theory defined by the above action is known as the non-abelian Toda
theory \REF\LS{A.N. Leznov and M.V. Saveliev, {\it Two-dimensional exactly and
completely integrable dynamical systems},
\CMP 89 1983 59 .} [\LS]
associated with the Lie algebra $B_2$
\REF\GS{J.-L. Gervais and M.V. Saveliev, {\it Black holes from non-abelian
Toda theories}, \PL B286 1992 271 .}
 [\GS].
The general solution of the
equations of motion is in principle contained in ref. \GS\ where it is shown
how the solutions for an equivalent system of equations can be obtained from
the general scheme of ref. \LS. However, it is non-trivial to actually spell
out the solution and put it in a compact and useful form. This was done
in
\REF\NAT{A. Bilal, {\it Non-abelian Toda theory: a completely integrable
model for strings on a black hole background}, \NP B422 1994 258 .}
ref. \NAT.

The \eoms obtained from the action \uiii\ read
$$\eqalign{
\du\dv r&={\sh r\over \ch^3r}\du t\dv t +\sh 2r\, e^{2\f}\cr
\du\dv t&=-{1\over \sh r\, \ch r}\left( \du r\dv t+\du t\dv r\right)\cr
\du\dv\f&=\ch 2r\, e^{2\f} \ .}
\eqn\di$$
Using these equations of motion it is completely straightforward to show
that the following three quantities are conserved [\GS]:
$$\eqalign{
T\equiv T_{++}&=(\du r)^2+\th^2r\, (\du t)^2 +(\du \f)^2-\du^2\f\cr
V^\pm\equiv
V^\pm_{++}&={1\over \sqrt{2}} \left( 2\du\f-\du\right)
\left[e^{\pm i\n}\left(\du r\pm i \th r\, \du t\right)\right]\cr }
\eqn\dii$$
i.e.
$$\dv T=\dv V^\pm=0\ .
\eqn\diii$$
Here $\n$ is defined by
$$\dv\n=\ch^{-2}r\, \dv t\quad , \qquad \du\n=(1+\th^2r)\du t
\eqn\div$$
where the integrability condition is fulfilled due to the \eoms \di.

\section{The symplectic structure and the constraint algebra}

Now I will define the symplectic structure using canonical Poisson brackets.
Recall that $u=\t+\s,\ v=\t-\s$. The action  \uiii\ can then equivalently
be written as (as usual, a dot denotes $\d_\t$
while a prime denotes $\d_\s$)
$$ S= {1\over \gd}\int \rd\t \rd\s
\left[ {1\over 2} (\dot r^2-r'^2)+{1\over 2}\th^2 r\, (\dot t^2-t'^2)
+{1\over 2} (\dot \f^2-\f'^2) + 2\, \ch 2r\, e^{2\f} \right] \ .
\eqn\ti$$
The constant $\gd$ can be viewed as the Planck constant $2\pi \hbar$,
already included ino the classical action, or merely as a coupling
constant. The canonical momenta then are
$$\pr=\gmd\, \dot r\quad , \quad \pt=\gmd \th^2 r\, \dot t\quad , \quad
\pf=\gmd\, \dot\f
\eqn\tii$$
and the canonical (equal $\t$) Poisson brackets are
$$\eqalign{
\{r(\t,\s)\, ,\, \pr(\t,\s')\} &=
\{t(\t,\s)\, ,\, \pt(\t,\s')\} =
\{\f(\t,\s)\, ,\, \pf(\t,\s')\} = \ds\cr
\{r(\t,\s)\, ,\, r(\t,\s')\} &=
\{r(\t,\s)\, ,\, t(\t,\s')\} = \ldots = 0\cr
\{\Pi_i(\t,\s)\, ,\, \Pi_j(\t,\s')\}&=0\ . \cr
}
\eqn\tiii$$
It follows that
the only non-zero equal $\t$ Poisson brackets are
$$\eqalign{
\{r(\t,\s)\, ,\, \dot r(\t,\s')\} =\gd \ds\quad &, \quad
\{t(\t,\s)\, ,\, \dot t(\t,\s')\} ={\gd \over \th^2 r}\ds\ ,\cr
\{\f(\t,\s)\, ,\, \dot \f(\t,\s')\} =\gd \ds\quad &, \quad
\{\dot r(\t,\s)\, ,\, \dot t(\t,\s')\} ={2\gd \over \sh r\, \ch r}\, \dot t
\, \ds\ ,\cr  }
\eqn\tiv$$
and those derived from them by applying $\d_\s^n \d_{\s'}^m$.

Before one can compute the Poisson bracket algebra of the $T, V^\pm$ one
has to rewrite them in terms of the fields and their momenta.  This means
in particular that second (and higher) $\t$-derivatives have to be
eliminated first, using the equations of motion. One might object that one
is not allowed to use the \eoms in a canonical formulation. However, the
conserved quantities given above are only defined up to
terms that vanish on solutions of the equations of motion. So the correct
starting point for a canonical formulation are the expression where all
higher $\t$-derivatives are eliminated, while the expressions given in
eq. \dii\ are merely derived from the canonical ones by use of the
equations of motion.
One has, for example,  using the $\f$-equation of motion \di\
$$\du^2\f={1\over 4}(\ddot\f+2\dot\f'+\f'')={1\over 2}(\f''+\dot\f')+\ch 2r
\, e^{2\f}=\d_\s\du\f+\ch 2r\,  e^{2\f}
\eqn\tv$$
where in the canonical formalism
$$\du r= {1\over 2}(\gd \pr+ r')\ ,\
\du t= {1\over 2}(\gd \th^{-2} r\,  \pt+ t')\ ,\
\du \f= {1\over 2}(\gd \pf+ \f')\ .
\eqn\tva$$
The canonical expressions for the $++$ components of the constraints are
$$\eqalign{
T&=(\du r)^2+\th^2 r\, (\du t)^2 + (\du\f)^2 -(\du\f)' -\ch 2r
e^{2\f}\cr
V^\pm&={1\over \sqrt{2}} e^{\pm i\nu}\Big[ 2\du\f\du r\pm 2i \th r\du\f\du t
+  2\th^3 r(\du t)^2 \mp 2i \th^2 r\du r\du t\cr
&\phantom{=e^{\pm i\nu}\Big[}+{\sh r\over \ch^3 r}\du t\, t'\mp i {\du
r\, t'+\du t\, r'\over \ch^2 r}-(\du r)'\mp i\th r(\du t)'-\sh 2r\,
e^{2\f}\Big]\ \cr }
\eqn\tvb$$
where the substitutions \tva\ are understood.
For the Poisson bracket of $T$ with itself one then obtains
$$\gmd \{T(\s)\, ,\, T(\s')\}  =
(\d_\s-\d_{\s'})\left[ T(\s') \ds\right]-{1\over 2} \dsppp \ .
\eqn\tvii$$
$\Tb$ satisfies the same algebra (with $\s\to -\s, \s'\to  -\s'$)
while
$\{T(\s)\, ,\, \Tb(\s')\}  =0$.
These are just two copies of the conformal algebra. If $\s$
takes values on the unit circle one can define the modes
$$
L_n=\gmd \int_{-\pi}^{\pi}\rd\s\left[ T(\t,\s)+{1\over 4}\right]
e^{in(\t+\s)}
\eqn\tx$$
and similarly for $\bar L_n$.
Then the bracket \tvii\ becomes a Virasoro algebra
$$
i\{L_n\, ,\, L_m\} =(n-m)L_{n+m}+{c\over 12} (n^3-n)\delta_{n+m,0}
\eqn\txi$$
and idem for $\{\bar L_n\, ,\, \bar L_m\}$,
while $\{L_n\, ,\, \bar L_m\}=0$. Here $c$ is the central charge given by
$$c={12 \pi\over \gd}\ .
\eqn\txii$$
The occurrence of a central charge already at the classical, Poisson bracket
level is due to the $\d^2\f$-term in $T$. This is reminiscent of the
well-known Liouville theory. The factor $i$ on the left hand side of
equations \txi\ may seem strange at first sight, but one should remember
that  quantization replaces $i$ times the canonical Poisson bracket by the
commutator. Hence \txi\ is indeed the Poisson bracket version of the
(commutator) Virasoro algebra.

In order to compute Poisson brackets involving $V^\pm$ one needs the
Poisson brackets involving the field $\nu$. Now, $\nu$ is only defined
through its partial derivatives, and thus only up to a constant. This
constant, however, may have a non-trivial Poisson bracket with certain modes
of $\nu$ and/or of the other fields. From \div\ one has using \tii
$$\nu'=\gd\pt+t'
\eqn\txiii$$
whereas one does not need $\dot\nu$ explicitly. Equation \txiii\ implies
$$
\d_\s\d_{\s'}\{\nu(\s)\, ,\, \nu(\s')\}=\{\nu'(\s)\, ,\, \nu'(\s')\} = 2\gd
\dsp\ .
\eqn\txiv$$
This is integrated to yield $\{\nu(\s)\, ,\, \nu(\s')\}=-\gd \es +h(\s)-
h(\s')$ where I already used the antisymmetry of the Poisson bracket.
$\es$ is defined to be $+1$ if $\s>\s'$, $-1$ if $\s<\s'$ and $0$ if
$\s=\s'$. The freedom to choose the function $h$ corresponds to the
above-mentioned freedom to add a constant $\nu_0$ to $\nu$ with $\{\nu(\s)\,
,\, \nu_0\}=h(\s)$. However, if one imposes  invariance under translations
$\s\to\s+ a,\ \s'\to\s' +a$ then $h$ can be only linear and one arrives at
$$\{\nu(\s)\, ,\, \nu(\s')\}=-\gd \e_\a(\s-\s')\equiv -\gd\left[
\es-{\a\over\pi} (\s-\s')\right]\ .
\eqn\txv$$
There is only one parameter $\a$ left, related to the zero-mode of $\nu$.

Before doing the actual calculation it is helpful to show how the result is
constrained by dimensional and symmetry considerations. First consider
$\{V^+(\s)\, ,\, V^+(\s')\}$. Each $V^+$ contains a factor
$e^{i\nu}$. The Poisson bracket of $e^{i\nu(\s)}$ with $e^{i\nu(\s')}$
leads to a term $\gd \es V^+(\s) V^+(\s')$. All other terms are
local, i.e. involve $\ds$ or derivatives of $\ds$. On dimensional
grounds\foot{
The naive dimensional counting assigns dimension 1 to each derivative,
dimension 0 to all fields $r,t,\f, \nu$ and functions thereof, except for
functions of $\f$. $e^{2\f}$ has dimension 2 as seen from the action,
while $\ds$ has dimension 1.}
$\ds$ must be multiplied by a dimension 3 object (3 derivatives) and $\dsp$
by a dimension 2 object. Furthermore, these objects must contain an
overall factor $e^{2i\nu}$. If the $T, V^+$ and $V^-$ form a closed
algebra, there are no such objects. The same reasoning applies to
$\{V^-(\s)\, ,\, V^-(\s')\}$. Thus one expects
$$\gmd\{V^\pm(\s)\, ,\, V^\pm(\s')\}=\es
V^\pm(\s)V^\pm(\s')\ .
\eqn\txvii$$
It is a bit tedious, but otherwise straightforward to verify that this is
indeed correct. Note that it is enough to do the computation for $V^+$
since $V^-$ is the complex conjugate of $V^+$ (treating the fields $r,t,\f$
and $\nu$ and their derivatives as real):
$$V^-=(V^+)^*\ .
\eqn\txviii$$

What can one say about $\{V^+(\s)\, ,\, V^-(\s')\}$? Using
\txviii\ one sees that $\{V^+(\s)\, ,\, V^-(\s')\}^* =
-\{V^+(\s)\, ,\, V^-(\s')\}\vert_{\s\leftrightarrow \s'}$. This
fact, together with the same type of arguments as used above, implies
$$\eqalign{
\gmd \{V^+(\s)\, ,\, V^-(\s')\}=&-\es
V^+(\s)V^-(\s') +(\d_\s-\d_{\s'})[a\ds] \cr
&+i b \ds +i(\d_\s^2+\d_{\s'}^2)[d\ds] +\tilde c \dsppp\cr
}
\eqn\txix$$
where $a,b,d,\tilde c$ are real and have (naive) dimensions 2,3,1 and 0.
Hence $\tilde c$ is a c-number. Also, $b,a,d$ cannot contain a factor
$e^{\pm i \nu}$. If one assumes closure of the algebra one must have $b=d=0$
and $a\sim T_{++}$. After a really lengthy computation one indeed finds
equation \txix\ with $b=d=0$, $a=T_{++}$ and $\tilde c=-{1\over 2}$.

Finally the Poisson bracket of $T$ with $V^\pm$ simply shows that
$V^\pm$ are conformally primary fields of weight (conformal dimension)
2. The complete algebra thus is
$$\eqalign{
\gmd \{T(\s)\, ,\, T(\s')\}  &=
(\d_\s-\d_{\s'})\left[ T(\s') \ds\right]-{1\over 2} \dsppp \cr
\gmd \{T(\s)\, ,\, V^\pm\s')\}  &=
(\d_\s-\d_{\s'})\left[ V^\pm(\s') \ds\right]\cr
\gmd \{V^\pm(\s)\, ,\, V^\pm(\s')\}&=\es
V^\pm(\s)V^\pm(\s')\cr
\gmd \{V^\pm(\s)\, ,\, V^\mp(\s')\}&=-\es
V^\pm(\s)V^\mp(\s')\cr
&\phantom{=}+(\d_\s-\d_{\s'})\left[ T(\s') \ds\right] -{1\over 2}
\dsppp \ .\cr }
\eqn\txx$$
The algebra of the $--$ components $\Tb, \Vb^\pm$
looks exactly the same except for the replacements $\s\to
-\s,\ \s'\to -\s'$ and hence $\d\to -\d,\ \es\to -\es$.

The algebra  \txx\ is the correct algebra for $\s\in {\bf R}$. If $\s\in
S^1$,  one must replace $\es\to \e_1(\s-\s')$ which is a
periodic function (cf. eq. \txv). Also $\ds\to {1\over 2}\d_\s
\e_1(\s-\s')=\ds-{1\over
2\pi}$ while $\dsp$ remains unchanged. But since the right hand sides of
\txx\ can be written using only $\es,\, \dsp$ and $\dsppp$, only the
replacement $\es\to \e_1(\s-\s')$ is relevant. One then defines the modes
$$
V^\pm_n= \gmd \int_{-\pi}^\pi \rd\s V^\pm(\t,\s)\, e^{in(\t+\s)}
\eqn\txxi$$
and similarly for $\bar V^\pm_n$.
The mode algebra is\foot{As a further consistency check one can verify that
the Jacobi identities are satisfied.}
$$\eqalign{
i\{L_n\, ,\, L_m\} &=(n-m)L_{n+m}+{c\over 12} (n^3-n)\delta_{n+m,0}\cr
i\{L_n\, ,\, V^\pm_m\} &=(n-m)V^\pm_{n+m}\cr
i\{V^\pm_n\, ,\, V^\pm_m\} &={12\over c}\sum_{k\ne 0} {1\over k}
V^\pm_{n+k}V^\pm_{m-k}\cr
i\{V^\pm_n\, ,\, V^\mp_m\} &=- {12\over c}\sum_{k\ne 0} {1\over k}
V^\pm_{n+k}V^\mp_{m-k}+ (n-m)L_{n+m}+{c\over 12} (n^3-n)\delta_{n+m,0}\ .\cr
}
\eqn\txxii$$
This is a non-linear algebra, reminiscent of the $W$-algebras. What is new
here are the non-local terms involving $\es$, or the ${1\over k}$ in the
mode algebra.

\section{The classical solutions of the \eoms }

The essential point for solving the \eoms \di\ is to realize the underlying
Lie algebraic structure. Following Gervais and Saveliev
[\GS], one first introduces fields $\au, \ad, \ap$ and $\am$
subject to the following \eoms
$$
\eqalign{
&\du\dv \au=-2(1+2\ap\am)e^{-\au}\cr
&\du\dv(2\ad-\au)+2\du(\dv\ap \am)=0\cr
&\du\left(e^{\au-2\ad}\dv\ap\right)=2\ap e^{-2\ad}\cr
&\du\left[e^{-\au+2\ad}\left(\dv\am-\am^2\dv\ap\right)\right]=2 \am
(1+\ap\am) e^{-2\au+2\ad}\ .\cr }
\eqn\dvii$$
It is then straightforward, although a bit lengthy, to show that we can
identify
$$\eqalign{
\f&=-{1\over 2}\au\cr
\sh^2 r&=\ap\am\cr
\du t&={1\over 2i}\left[(1+2\ap\am){\du\am\over \am}- {\du\ap\over \ap}
-2(1+\ap\am)\du(\au-2\ad)\right]\cr
\dv t&={i\over 2}\left[(1+2\ap\am){\dv\ap\over \ap}- {\dv\am\over \am}
\right]\cr}
\eqn\dviii$$
i.e. with these definitions  the equations \dvii\ and \di\ are equivalent.
The integrability condition for solving for $t$ is given by the
equations \dvii.

The advantage of equations \dvii\ over equations \di\ is that they follow
 from a Lie algebraic formulation. The relevant algebra is $B_2$ with
generators $h_1, h_2$ (Cartan subalgebra) and $E_{e_1}, E_{e_2}, E_{e_1-e_2},
E_{e_1+e_2}$ and their conjugates $E_\a^+=E_{-\a}$. Then $H=2h_1+h_2,
J_+=E_{e_1}$ and $J_-=E_{-e_1}$ span an $A_1$ subalgebra. $H$ induces a
gradation on $B_2$. The gradation 0 part ${\cal G}_0$  is spanned by $h_1,
h_2, E_{e_2}$ and $E_{-e_2}=E_{e_2}^+$. The corresponding group elements
$g_0\in G_0$ can be parametrized as
$$g_0=\exp(\ap E_{e_2})\exp(\am E_{e_2}^+)\exp(\au h_1 +\ad h_2)\ .
\eqn\dx$$
One can then show that equations \dvii\ are equivalent to
$$\du( g_0^{-1}\dv g_0)=[J_-\, ,\, g_0^{-1}J_+ g_0]\ .
\eqn\dix$$
If one now defines $g_+$ and $g_-$ as solutions of the ordinary differential
equations
$$\eqalign{
\du g_+^{-1}&=g_+^{-1} (g_0^{-1} J_+ g_0)\cr
\dv g_-&=g_- (g_0 J_- g_0^{-1})
\ .\cr}
\eqn\dxii$$
and $g=g_-g_0g_+$ then one can show [\GS,\NAT] that eq. \dix\ is in turn
equivalent to
$$\du (g^{-1}\dv g)=0 \quad {\rm and} \quad  \dv (\du g g^{-1})=0 \ .
\eqn\dxiv$$
The general solution to these equations is well-known:
$g=g_L(u) g_R(v)$.  But each group element $g_L$ and $g_R$ has again a Gauss
decomposition $g_L(u)=g_{L-}(u)g_{L0}(u)g_{L+}(u)$ and
$g_R(v)=g_{R-}(v)g_{R0}(v)g_{R+}(v)$ so that
$g=g_{L-}(u)g_{L0}(u)g_{L+}(u)g_{R-}(v)g_{R0}(v)g_{R+}(v)$.
On the other hand we also have the decomposition $g=g_-g_0g_+$ where $g_+$ and
$g_-$
must obey the differential
equations \dxii.
The latter translate into
$$\eqalign{
\du g_{L+}(u)=-{\cal F}_L(u) g_{L+}(u) \quad &,
\quad {\cal F}_L(u) =g_{L0}^{-1}(u) J_+ g_{L0}(u)\ ,\cr
\dv g_{R-}(v)= g_{R-}(v){\cal F}_R(v) \quad &,
\quad {\cal F}_R(v) =g_{R0}(v) J_- g_{R0}^{-1}(v)\ .\cr
}
\eqn\dxx$$

The strategy then is
\pointbegin
Pick some arbitrary $g_{L0}(u),\ g_{R0}(v)\in G_0$.
\point
Compute the solutions $g_{L+}(u)$ and $g_{R-}(v)$ from the first order
ordinary differential equations \dxx.
\point
Let
$$
\Gamma=g_{L0}(u) g_{L+}(u) g_{R-}(v) g_{R0}(v)
\eqn\dxxi$$
and choose a basis $\la$ of states annihilated by $\G_+$ (i.e. by $E_{e_1},
 E_{e_1-e_2}$ and $E_{e_1+e_2}$. Then using the different decompositions of $g$
 we have
$$G_{\a\b}\equiv \lb g_0 \la=\lb \Gamma \la\ .
\eqn\dxxii$$
This yields all matrix elements of $g_0$, solution of equation \dix, which
in turn, as shown above, yields the solution for the $\au,\ad,\ap$ and
$\am$, parametrized in terms of the arbitrary $g_{L0}(u)$ and $g_{R0}(v)$.
\par

I will now display the resulting solutions. For details of the derivation, see
ref. \NAT. One introduces three arbitrary (``left-moving") functions $f_1(u),
f_+(u)$ and $f_-(u)$ of {\it one} variable (parametrizing  $g_{L0}(u)$),
and  three arbitrary
(``right-moving") functions $g_1(v),
g_+(v)$ and $g_-(v)$ of {\it one} variable (parametrizing $g_{R0}(v)$).
To write the results in a more compact way,
introduce the functions of one variable
$$\eqalign{
F_1(u)&=-\int^u e^{f_1} (1+2f_+f_-)\cr
F_2(u)&=2\int^u e^{f_1} f_-\cr
F_3(u)&=-2\int^u e^{f_1}f_+(1+f_+f_-) \cr}
\eqn\dxxxvi$$
as well as
$$F_+=F_1+f_+ F_2\quad , \quad F_-=F_3-f_+ F_1
\eqn\dxxxvii$$
and similarly for $G_1(v)$ etc, with $f_i(u)\to g_i(v)$. Then one introduces
the quantities $X,Y,Z$ and $V,W$ that depend on both variables $u$ and $v$:
$$\eqalign{
X&=1+f_+g_++F_+G_++F_-G_-\cr
Y&=(1+f_+f_-)(1+g_+g_-)+f_-g_-+(F_1-f_-F_-)(G_1-g_-G_-)\cr
&\phantom{=}+(F_2+f_-F_+)(G_2+g_-G_-)\cr
Z&=1+2F_1G_1  +F_2G_2+F_3G_3+(F_1^2+F_2F_3)(G_1^2+G_2G_3)\cr
V&=-g_--f_+-g_+g_-f_+-g_-F_+G_+-g_-F_-G_-+F_-G_1-F_+G_2\cr
W&=-f_--g_+-f_+f_-g_+-f_-F_+G_+-f_-F_-G_-+F_1G_--F_2G_+\ .\cr
}
\eqn\dxxxxiv$$
They are sums of products of left-moving times right-moving quantities.
 The complete solution then is
$$\eqalign{
e^{a_1}&= e^{-f_1-g_1}Z\cr
e^{-a_2}&=\pm e^{f_2+g_2}{Y\over Z}\cr
\ap&=e^{f_1-2f_2}{V\over Y}\cr
\am&=e^{2f_2-f_1}{YW\over Z}\cr
}
\eqn\dxxxxvii$$
while the fifth equation is the relation
$XY-Z=VW$.
{}From equations \dviii\ one immediately finds
$$\eqalign{
\f&={1\over 2}\left( f_1+g_1-\log Z\right)\cr
\sh^2 r&={VW\over Z}={XY\over Z}-1\cr
t&=t_0+i\int^u f_-f_+'-i\int^v g_-g_+' +{i\over 2} \log {V\over W} \ .
\cr }
\eqn\dxxxxxiii$$
which is the general solution of the \eoms.

Using the \eoms for $\f,\, r$ and $t$ it was shown above that the
quantities $T\equiv T_{++}$ and $V^\pm\equiv V^\pm_{++}$ are conserved, i.e.
can only depend on $u$. This means that they must be expressible entirely in
terms of the $f_i(u)$'s. Given the complexity of the solutions
\dxxxxxiii\ and \dxxxxiv\ this is highly  non-trivial and
constitutes a severe consistency check. The same considerations
apply to $\Tb\equiv T_{--}$ and $\Vb^\pm\equiv V^\pm_{--}$.
One can indeed show [\NAT] that
$$\eqalign{
T_{++}&={1\over 4}(f_1')^2-{1\over 2}f_1''+  f_+'(f_-'-f_-^2f_+') \cr
V^+&={1\over \sqrt{2} } (f_1'-\du) \, \left[
\exp\left( -2\int^u f_-f_+'\right)  (f_-'-f_-^2f_+') \right]\cr
V^-&={1\over \sqrt{2} } (f_1'-\du) \, \left[
\exp\left( +2\int^u f_-f_+'\right)  f_+' \right]\ . \cr}
\eqn\ddxiii$$
It is then natural to set
$$\eqalign{
f_-&=e^{\sqrt{2}\vf_1}\cr
f_+'&={1\over \sqrt{2}}e^{-\sqrt{2}\vf_1} (\d_\s\vf_1+i\d_\s\vf_2)\cr
f_1&=\sqrt{2}\vf_3\cr
}
\eqn\txxxviii$$
(and with a similar relation between the $g_i$ and $\bar\vf_i$).
The conserved quantities $T, V^\pm$ are easily expressed in terms of the
$\vf_j$ as
$$\eqalign{
T&={1\over 2} \sum_{j=1}^3 (\d_\s \vf_j)^2 -{1\over \sqrt{2}}\d_\s^2\vf_3\cr
V^\pm&={1\over 2}(\sqrt{2}\d_\s\vf_3-\d_\s)\left[ e^{\mp i\sqrt{2}\vf_2}
(\d_\s\vf_1\mp i\d_\s\vf_2)\right]\ .\cr
}
\eqn\txxxxiii$$
Thus in terms of the $\vf_j$, $T$ has the standard form of a stress-energy
tensor
in a conformal field theory with a background charge.  The $V^\pm$ are
{\it local} expressions of the
fields $\vf_j$, analogous to standard vertex operators. (Of course, their
Poisson
brackets exhibit the non-local $\es$-function.)

\section{Canonical transformation to free fields}

In principle one could now deduce the Poisson brackets
of the $f_i$ and $g_i$ through the transformation induced by the classical
solution \dxxxxxiii\ and \txxxviii.
More precisely, one would have to allow formally that the $f_i$ and $g_i$,
respectively the $\vf_i$ and $\bar\vf_i$,
depend both on $u$ and $v$ since one has to consider the full phase space
and not only the manifold of solutions to the equations of motion.
Nevertheless, equations \dxxxxxiii\ and  \txxxviii, as well as their time
derivatives, constitute a phase space transformation from $r(\t,\s),\,
t(\t,\s),\, \f(\t,\s)$ and their momenta $\pr(\t,\s),\, \pt(\t,\s),\,
\pf(\t,\s)$ to new phase space variables $f_i(\t,\s),\ g_i(\t,\s)$ and to
$\vf_i(\t,\s), \bar\vf_i(\t,\s)$. (Of
course, the equations of motion still imply $\dv f_i=\du
g_i=\d_v\vf_i=\d_u\bar\vf_i=0$.) In
practice, this would be very complicated to implement. It is much simpler
to use the Poisson brackets of the $T$ and $V^\pm$ derived before, and then
consider the $T$ and $V^\pm$ (or $\Tb$ and $\Vb^\pm$) as given in terms of
the $\vf_i$ only (or $\bar\vf_i$ only).  Thus one does the phase space
transformation in two steps: $r(\t,\s),\,
t(\t,\s),\, \f(\t,\s),\, \pr(\t,\s),\, \pt(\t,\s),\,
\pf(\t,\s)\ \to\ T_{\pm\pm}(\t,\s),\,    V^+_{\pm\pm}(\t,\s),\,
V^-_{\pm\pm}(\t,\s)\    \to\  \vf_i(\t,\s),\, \bar\vf_i(\t,\s)$.
This yields the Poisson brackets of the $\vf_i$ and of the $\bar\vf_i$ in a
relatively easy way. These Poisson brackets are simple, standard
harmonic oscillator Poisson brackets:
$$\{\d_\s \vf_i(\s)\, ,\, \d_{\s'} \vf_j(\s')\}=\gd \delta_{ij} \dsp
\eqn\txxxix$$
or
$$\{\vf_i(\s)\, ,\, \vf_j(\s')\}=-{\gd\over 2} \delta_{ij} \es\ .
\eqn\txxxx$$
If one considers $\s\in S^1$, the mode expansion\foot{
Note that this is a Fourier expansion in $\s$. The factor $e^{-in\t}$ is
only extracted from the $\vf_n^j$ for convenience. The $\vf_n^j$ are still
functions of $\t$. It is only if one imposes the equations of motion that
the $\vf_n^j$ are constant.}
$$\d_\s \vf_j(\t,\s)={\g\over \sqrt{2\pi}} \sum_n
\vf^j_n e^{-in(\t+\s)} \eqn\txxxxi$$ %
leads to
$$i\{\vf^j_n\, ,\, \vf^k_m\}=n \delta^{ij} \delta_{n+m,0}
\eqn\txxxxii$$
as appropriate for three sets of harmonic oscillators. The Poisson brackets of
the
$\bar\vf_i$ are analogous, while $\{\vf_i(\s)\, ,\, \bar\vf_j(\s')\}=0$.

\section{Associated linear differential equation}

{}From experience with integrable models, in particular the (non-affine,
conformally invariant) Toda models
\REF\BG{A. Bilal and J.-L. Gervais, {\it Systematic approach to
conformal systems with extended Virasoro symmetries}, Phys. Lett.
{\bf B206} (1988) 412; {\it Systematic construction of $c=\infty$
conformal systems from classical Toda field theories},  Nucl. Phys.
{\bf B314} (1989) 646; {\it Systematic construction of conformal
theories with higher-spin Virasoro symmetries}, Nucl. Phys. {\bf B318}
(1989) 579.}
[\BG] one expects that the conserved quantities appear as coefficients of
an ordinary linear differential equation, e.g. for the $A_{m-1}$ Toda model
$$\left[\du^m-\sum_{k=2}^m W^{(k)}(u) \du^{m-k}\right]\psi(u) =0\ .
\eqn\qi$$
The $ W^{(k)}(u), k=2,\ldots m$ are the conserved quantities which form the
$W_m$-algebra, while the solutions $\psi_j(u)$ of this equation, together
with solutions $\chi_j(v)$ of a similar equation in $v$, are the
building blocks of the general solution to the Toda equations of motion.
The $ W^{(k)}(u)$ have (naive) dimension $k$.

In the present theory all conserved quantities have dimension 2,
so one might expect a second-order  differential
equation of the form $(-\d^2+U)\psi=0$. To fit three conserved quantities into
$U$, it needs to be at least a $2\times 2$-matrix. In ref. \NAT, I guessed the
following
linear differential equation
$$\left[ \d_u^2-\pmatrix{ \a T(u)& \b_+ V^+(u)\cr \b_- V^-(u) & \delta T(u)
\cr } \right] \Psi(u)=0\ .
\eqn\qii$$

Actually, if one inserts the free-field representation \txxxxiii\ of $T$ and
$V^\pm$ one
can see that this equation admits the very simple solution
$$\eqalign{
\psi_1&=\exp(a\vf_1 +ib\vf_2 +d\vf_3)\cr
\psi_2&=\exp(a\vf_1 -ib\vf_2 +d\vf_3)\ .\cr
}
\eqn\qqi$$
if and only if
$$\eqalign{
a={1\over \sqrt{2}}\quad &, \quad b=d= -{1\over \sqrt{2}}\cr
\a=\delta=1 \quad &, \b_+=\b_-=-\sqrt{2}\ .\cr
}
\eqn\qqii$$
The existence of this very simple solution already is an indication that the
differential equation \qii\ is probably the correct generalization of \qi\
for $W$-algebras to the present $V$-algebra. That this is indeed so was shown
in
\REF\MCK{A. Bilal, {\it Multi-component KdV hierarchy, $V$-algebra and
non-abelian Toda theory}, Lett. Math. Phys. {\bf 32} (1994) 103.}
ref. \MCK, and will be reviewed in the next section.

\chapter{The $V$-algebra as second Gelfand-Dikii bracket,\break
 the resolvent and matrix KdV-hierarchy}

\def\d{\partial}
\def\dsi{\partial_\sigma}

\def\du{\partial_u}
\def\dv{\partial_v}
\def\f{\varphi}

\def\s{\sigma}
\def\sp{\sigma'}
\def\l{\lambda}
\def\t{\tau}
\def\is{\int {\rm d} \sigma\, }

\def\ds{\delta(\s-\s')}
\def\dsp{\delta'(\s-\s')}

\def\dsppp{\delta'''(\s-\s')}
\def\es{\epsilon(\s-\s')}

\def\a{\alpha}
\def\b{\beta}
\def\g{\gamma}
\def\gd{\gamma^2}
\def\gmd{\gamma^{-2}}
\def\rmd{{\rm d}}
\def\rd{\sqrt{2}}
\def\la{\langle}
\def\ra{\rangle}

\def\vp{{V^+}}
\def\vm{{V^-}}
\def\GD{Gelfand-Dikii\ }
\def\P{\Psi}

\def\dd #1 #2{{\delta #1\over \delta #2}}
\def\tr{{\rm tr}\ }

\REF\BAK{I. Bakas, {\it Higher spin fields and the \GD algebras},
Commun. Math. Phys. {\bf 123} (1989) 627.}
In this section, following ref. \MCK, I will relate the $V$-algebra obtained
above
 to well-known
mathematical structures. In the case of the standard $W_m$-algebras it was
shown [\BAK]
 that their classical version coincides with second Gelfand-Dikii symplectic
structure
\REF\GELDA{I.M. Gel'fand and L.A. Dikii, {\it Fractional powers of
operators and hamiltonian systems},
 Funct. Anal. Applic. {\bf 10} (1976) 259;
 {\it The resolvent and
hamiltonian systems}, Funct. Anal. Applic. {\bf 11} (1977) 93.}
\REF\ADLER{M. Adler, {\it On a trace functional for
pseudo-differential operators and the symplectic structure of the
Korteweg-de Vries type equations}, Invent. math. {\bf 50} (1979)
219.}
\REF\LEBMAN{D.R. Lebedev and Yu.I. Manin, {\it Gel'fand-Dikii
Hamiltonian operator and the coadjoint representation of the Volterra
group}, Funct. Anal. Applic. {\bf 13} (1979) 268.}
\REF\GELDB{I.M. Gel'fand and L.A. Dikii, {\it A family of
Hamiltonian structures connected with integrable non-linear
differential equations}, Inst. Appl. Math. Acad. Sci. USSR preprint
n$^{\rm o}$ 136, 1978 (in Russian).}
 \REF\KW{B.A. Kupershmidt and G. Wilson, {\it Modifying Lax equations
and the second Hamiltonian structure}, Invent. Math. {\bf 62} (1981)
403.}
\REF\DIK{L.A. Dikii, {\it A short proof of a Kupershmidt-Wilson
theorem}, Commun. Math. Phys. {\bf 87} (1982) 127.}
[\GELDA-\DIK] on the space of ordinary differential operators of order $m$ (as
in eq.
\qi). What I will show here in the remainder of this contribution,
 is that a straightforward matrix generalization
 leads to a whole family of $V$-algebras. In this section I will obtain
 the algebra of the previous section
 from the $2\times 2$-matrix second-order differential operator
$$ L=\d^2-U\ ,\quad
U=\pmatrix{ T& -\rd V^+\cr -\rd V^- & T \cr }\ .
\eqn\uiii$$
The general case of $n\times n$-matrix $m^{\rm th}$-order differential
operators
leading to $V_{n,m}$-algebras will be treated in the next section.

If not indicated otherwise, $\d\equiv \dsi$, and $U$
depends on $\s$. $U$ may also depend on other parameters $t_1, t_2, \ldots$.
In the previous section $U$ depended on $\t$ and $\s$, but, upon imposition of
the
\eoms, only through
 the combination $\s+\t$. Actually, as usual (see below),
this is the first flow of the matrix KdV hierarchy: ${\d\over \d t_1}
U=\dsi U$, so that $\t$ is identified with $t_1$.

Let $f$ and $g$ be differential polynomial functionals on the space of
second-order differential operators $L$, i.e. polynomial functionals
of $U$ (and its derivatives). One defines the pseudo-differential
operator %
$$X_f=\d^{-1} X_1 +\d^{-2} X_2\quad , \quad X_1=\dd f U
\eqn\di$$
where\foot{
On the circle $S^1$ e.g., $\d^{-1}$ is well-defined on functions
without constant Fourier mode, i.e. $f(\s)=\sum_{m\ne 0}
f_m e^{-im\s}$. One easily sees that $(\d^{-1}f)(\s)=\int \rmd\sp\
{1\over 2} \es f(\sp)$ with $\es = {1\over \pi i}\sum_{m\ne 0} {1\over
m} e^{im(\s-\sp)}$.
}
 $\d^{-1}\d=\d\d^{-1}=1$ and ${\delta\over \delta U}$ is defined
as
$${\delta\over \delta U}=\pmatrix{
{1\over 2} {\delta\over \delta T}&
-{1\over \rd} {\delta\over \delta \vm}\cr
-{1\over \rd} {\delta\over \delta \vp}&
{1\over 2} {\delta\over \delta T}\cr }
\eqn\dii$$
so that ${\delta\over \delta U} \int \tr U^n =n U^{n-1}$, and $X_2$
is determined (cf. e.g. [\ADLER,\LEBMAN]) by requiring\foot{
This condition is necessary since the coefficient of $\d$ in $L$
vanishes. I will discuss this condition in more detail in the
 next section.} ${\rm res} [L,X_f]=0$. As usual, the residue of a
pseudo-differential operator, denoted ${\rm res}$, is the coefficient
of $\d^{-1}$. One then has
$$X_2'={1\over 2}\left( \dd f U \right)'' +{1\over 2}
\left[ U, \dd f U \right]\ .
\eqn\diii$$
Integrating this equation yields $X_2$. Here, one observes a new
feature as compared to the scalar case: since in general $\left[ U,
\dd f U \right]\ne 0$, $X_2$ will be given by a non-local
expression involving an integral. This is the origin of the
non-local terms ($\sim\es$) in the $V$-algebra \txx.

In analogy with
the standard procedure [\GELDA-\DIK], I define the second \GD bracket
in the matrix case as follows
$$\{f,g\}_{\rm GD2}=a\is\tr\ {\rm res} \left( L(X_f L)_+ X_g
-(L X_f)_+ L X_g \right) \ .
\eqn\dvi$$
Using the definitions of $L$ and $X_f, X_g$ it
 is straightforward to obtain
$$\eqalign{
\{f,g\}_{\rm GD2}=a\is\tr \Bigg( &{1\over 2} \dd f U \d^3 \dd g U
+{1\over 2} \left[ U, \dd g U \right] \left( \d^{-1}
\left[ U, \dd f U \right] \right)\cr
&-\dd f U (U\d+{1\over 2} U') \dd g U
+ \dd g U (U\d+{1\over 2} U') \dd f U
 \Bigg)\cr }
\eqn\dvii$$
where the $\d^{-1}$ is meant to act only on $\left[ U, \dd f U
\right]$.

Inserting the definitions of the $2\times 2$-matrices $U$ and ${\delta
\over \delta U}$ one obtains
$$\eqalign{
\{f,g\}_{\rm GD2}=-{a\over 2} \is \Bigg[ &
 -{1\over 2} \dd f T \d^3 \dd g T
 -{1\over 2} \dd f \vp \d^3 \dd g \vm
 -{1\over 2} \dd f \vm \d^3 \dd g \vp  \cr
&+T\left(\dd f T \d \dd g T + \dd f \vp \d \dd g \vm
 + \dd f \vm \d \dd g \vp -(f \leftrightarrow g) \right) \cr
&+\vp\left(\dd f T \d \dd g \vp + \dd f \vp \d \dd g T
 -(f \leftrightarrow g) \right) \cr
&+\vm\left(\dd f T \d \dd g \vm + \dd f \vm \d \dd g T
 -(f \leftrightarrow g) \right) \Bigg]\cr
 -{a\over 2} \int \int \rmd\s\rmd\sp
& \es\left( \vp \dd f \vp - \vm \dd f \vm \right)(\s)
\left( \vp \dd g \vp - \vm \dd g \vm \right)(\sp) \ .\cr }
\eqn\dix$$
For completeness, let me note that the first Gelfand-Dikii bracket
defined by
$\{f,g\}_{\rm GD1} = a \is\tr\   {\rm res} ([L,X_f]_+ X_g)$
 is simply
$$\{f,g\}_{\rm GD1}=-a\is\left( \dd f T \d \dd g T +
\dd f \vp \d \dd g \vm + \dd f \vm \d \dd g \vp \right) \ .
\eqn\dviii$$
Taking $f,g$ to be $T, \vp$ or $\vm$ one concludes that
the
second \GD bracket \dix\ coincides with the $V$-algebra \txx\
 provided
one chooses
$$ a=-2\gmd\ .
\eqn\dx$$
Henceforth I will adopt this choice and the only Poisson bracket used
is the second \GD bracket, unless otherwise stated.
Let me note that the free field representation of $T, V^\pm$ in terms of the
$\vf_i$
obtained in the previous section constitutes a Miura transformation since it
maps the
second \GD symplectic structure to the much simpler
 free-field Poisson brackets.  I will discuss the Miura transformation in more
detail
in the next section.

\REF\GELD{I.M. Gel'fand and L.A. Dikii, {\it Asymptotic behaviour of the
resolvent of Sturm-Liouville equations and the algebra of the Korteveg-de Vries
equations}, Russ. Math. Surv. {\bf 30}
(1975) 77.}
Let me now turn to some results that are a bit more specific to second-order
 differential operators, generalizing the classical work of ref. \GELD.
 For a
hermitian $n\times n$-matrix $U$ define the $n\times n$-matrix resolvent $R$ as
$$ R(x,y;\xi)=\la
x\vert (-\d^2+U+\xi)^{-1}\vert y\ra \eqn\cci$$ %
which is a solution of
$$(-\d_x^2 +U(x)+\xi)R(x,y;\xi)=\delta(x-y)\ .
\eqn\ccii$$
Just as in the scalar case, $n=1$, the restriction of the resolvent to the
diagonal, $R(x;\xi)\equiv R(x,x;\xi)$ has an asymptotic expansion for
$\xi\to\infty$ of the form
$$R(x;\xi)=\sum_{n=0}^\infty {R_n[u]\over \xi^{n+1/2}}\ .
\eqn\cciii$$
This equation is to be understood as an equality of the asymptotic
expansions in half-integer powers of $1/\xi$, disregarding any terms
that vanish exponentially fast as $\xi\to\infty$.
{}From the defining differential equation for $R(x,y;\xi)$ one easily
establishes that
 $R\equiv
R(x;\xi)=R(x,x;\xi)$ satisfies
$$R'''-2(UR'+R'U)-(U'R+RU')+[U,\d^{-1}[U,R]]=4\xi R'
\eqn\ccxaa$$
(where $R'\equiv \d_x R(x;\xi)$ etc.) and hence that the coefficients $R_n$ of
the asymptotic expansion \cciii\ satisfy
$$ 4R'_{n+1}=R_n'''-2(UR'_n+R'_nU)-(U'R_n+R_nU')+[U,\d^{-1}[U,R_n]]\ .
\eqn\ccxa$$
This allows us to determine the $R_n$ recursively:
$$\eqalign{
R_0=&{1\over 2}\cr
R_1=&-{1\over 4}U\cr
R_2=&{1\over 16}(3U^2-U'')\cr
R_3=&-{1\over 64}(10U^3-5UU''-5U''U-5{U'}^2+U^{(4)})\cr
R_4=&{1\over 256}(35U^4-21U^2U''-21U''U-28UU''U
                 -28{U'}^2U-28U{U'}^2-14U'UU'\cr
&\phantom{{1\over 256}(} +7UU^{(4)}+7U^{(4)}U+14U'U'''+14U'''U'
                 +21{U''}^2+U^{(6)})\ . \cr   }
\eqn\ccxb$$
\REF\OMG{E. Olmedilla, L. Martinez Alonso and F. Guil, {\it
Infinite-dimensional Hamiltonian systems associated with matrix
Schr\"odinger operators}, Nuovo Cim. {\bf 61B} (1981) 49.}
One of the reasons why one is interested in the coefficients $R_n$ is the
following:
if one defines an infinite family of Hamiltonians as
$$H_n={(-4)^n\over 2(2n-1)}\int \rmd x\ \tr R_n(x)
\eqn\cccia$$
one can show [\MCK] that all $H_n$ are in involution\foot{
Such a family of $H_n$ in involution for the matrix Schr\"odinger operator $L$
was already obtained a long time ago in ref.
\OMG.}:
$$\{ H_n,H_m\}=0\ .
\eqn\cccii$$
The proof uses the fact that the recursion relation between the $H_n$
(inherited from
the one between the $R_n$) relates the first and second \GD brackets:
$\{H_n,H_m\}_2\sim \{H_n,H_{m+1}\}_1=-\{H_{m+1},H_n\}_1\sim
-\{H_{m+1},H_{n-1}\}_2
= \{H_{n-1},H_{m+1}\}_2$, so
that by iteration one arrives at $\{H_n,H_m\}_2=\{H_0,H_{n+m}\}_2=0$.
The first few $H_n$ are
$$\eqalign{
H_1&={1\over 2}\int\tr\, U = \int T \cr
H_2&={1\over 2}\int\tr\, U^2 = \int (T^2+2\vp\vm) \cr
H_3&={1\over 2}\int\tr\, (2U^3+U'^2)
= \int (2T^3+12T\vp\vm+T'^2+2\vp'\vm') \cr
H_4&={1\over 2}\int\tr\, (5U^4+10UU'^2+U''^2)\cr
&= \int (5T^4+20\vp^2\vm^2+60T^2\vp\vm+10TT'^2\cr
&\phantom{= \int ( } +20T\vp'\vm'+20T'\vp\vm'+20T'\vp'\vm +T''^2
+2\vp''\vm'') \ .\cr}
\eqn\cccvi$$
Note that within our non-abelian Toda field theory,
 $H_1$ is the integral of the $(++)$-component of the energy-momentum
tensor $T\equiv T_{++}$. Thus the Hamiltonian of the theory is $H=L_0+\bar L_0=
H_1+\bar H_1$. The other $H_n$ are
all in involution with $H_1$ (and obviously also with $\bar H_1$)
 and thus are conserved under the conformal evolution of
the  non-abelian Toda field theory.

\REF\CD{F. Calogero and A. Degasperis, {\it Nonlinear evolution equations
solvable by the inverse spectral transform associated with the multichannel
Schr\"odinger problem, and properties of their solution}, Lett. Nuovo Cim. {\bf
15}
(1976) 65;
}
As usual one can define an infinite hierarchy of flows by\foot{
As before, $\{\cdot,\cdot\}$ is meant to be the second \GD bracket.
But since $4 {\d U\over \d t_r}=4\gmd \{U,H_r\}_{\rm GD2} =
\gmd \{U,H_{r+1}\}_{\rm GD1}$, one can also use the first \GD bracket
and the next higher Hamiltonian instead.}
$${\d U\over \d t_r}=\gmd \{U,H_r\}\ .
\eqn\si$$
Since $\{H_r,H_s\}=0$ it follows from the Jacobi identity that all
flows commute.

The flow in $t_2$ gives the matrix
generalisation of the KdV equation:\foot{
Matrix KdV flows were already discussed a long time ago in ref. \CD.}
$${\d U\over \d t_2}=(3U^2-U'')'
\eqn\sii$$
or in components
$${\d T\over \d t_2}=(3T^2-T''+6\vp\vm)'\quad , \quad
{\d V^\pm\over \d t_2}=(6TV^\pm-{V^\pm}'')'
\eqn\siii$$
Just as the Virasoro algebra is a subalgebra of the $V$-algebra \txx,
the KdV equation is simply obtained by setting $V^\pm=0$.
Note that since all $H_n$ are symmetric in $\vp$ and $\vm$ any
non-local term ($\sim\es$) that might appear cancels in all flow
equations, and the latter are always partial {\it differential}
equations.\foot{This also follows from the equivalence with the first
\GD bracket which is local, see above.}

\chapter{The \GD symplectic structure for general $n\times n$-matrix $m^{\rm
th}$-order differential operators and the $V_{n,m}$-algebras}

\def\d{\partial}
\def\dsi{\partial_\sigma}

\def\f{\varphi}
\def\s{\sigma}
\def\sp{\sigma'}
\def\l{\lambda}
\def\is{\int {\rm d} \sigma\, }

\def\ds{\delta(\s-\s')}
\def\dsp{\delta'(\s-\s')}

\def\dsppp{\delta'''(\s-\s')}
\def\es{\epsilon(\s-\s')}
\def\e{\epsilon}
\def\a{\alpha}
\def\b{\beta}
\def\g{\gamma}
\def\gd{\gamma^2}
\def\gmd{\gamma^{-2}}
\def\rmd{{\rm d}}
\def\rd{\sqrt{2}}
\def\vp{{V^+}}
\def\vm{{V^-}}
\def\GD{Gelfand-Dikii\ }
\def\P{\Psi}
\def\dd #1 #2{{\delta #1\over \delta #2}}
\def\bin #1 #2{{ #1\choose #2}}
\def\tr{{\rm tr}\ }
\def\res{{\rm res}\ }

\def\id{{\bf 1}}
\def\Lemma #1  {\noindent{\bf Lemma #1 :\ }}

\def\v{{\cal V}(f)}
\def\vt{{\tilde{\cal V}}(f)}
\def\ft{{\tilde f } }
\def\gt{{\tilde g } }
\def\P{{\cal P} }

\REF\MW{A. Bilal, {\it Non-local matrix generalizations of $W$-algebras},
Princeton
University preprint PUPT-1452 (March 1994), hep-th@xxx/9403197, Comm. Math.
Phys. in
press.}
In this section, following ref. \MW, I will compute the
 second \GD bracket of
two functionals $f$ and $g$ of the $n\times n$ matrix coefficient functions
$U_k(\s)$ of the linear $m^{\rm th}$-order differential operator\foot{
Throughout the rest of this paper, $m$ will denote the order of $L$ which is a
positive integer.}
$$L=-\d^m+\sum_{k=1}^mU_k\d^{m-k}\equiv \sum_{k=0}^mU_k\d^{m-k}\ .
\eqn\di$$
where $\d={d\over d\s}$.
To make subsequent formula more compact, I formally introduced
$U_0=-\id$. The fuctionals $f$ and $g$ one considers are of the form
$f=\int\tr P(U_k)$, where $P$ is some polynomial in the $U_k,\,
k=1,\ldots m$, and their derivatives (i.e. a differential polynomial
in the $U_k$). $P$ may also contain other constant or non-constant
numerical matrices so that these functionals are fairly general. (Under
suitable boundary conditions, {\it any} functional of the $U_k$ and their
derivatives can be approximated to arbitrary ``accuracy" by an $f$ of
the type considered.) The integral can either be defined in a formal
sense as assigning to any function an equivalence class by considering
functions only up to total derivatives (see e.g. section 1 of the second ref.
\GELDA), or in the standard way if one restricts the integrand, i.e. the
$U_k$, to the class of e.g. periodic functions or sufficiently fast
decreasing functions on ${\bf R}$, etc. All that matters is that the
integral of a total derivative vanishes and that one can freely
integrate by parts.

To define the \GD brackets, it is standard to use pseudo-differential
operators
[\ADLER,\LEBMAN] involving integer powers of $\d^{-1}$, as already encountered
in the
previous section. Again, $\d^{-1}$
can be defined in a formal sense by $\d \d^{-1}=\d^{-1}\d=1$, but one can
also give a concrete definition on appropriate classes of functions.
For example for $C^\infty$-functions $h$ on ${\bf R}$ decreasing
exponentially fast as $\s\to\pm\infty$ one can simply define
$(\d^{-1}h)(\s)=\int_{-\infty}^\infty \rmd \sp
{1\over 2}\es h(\sp)$.

Throughout this section, I will only state the results. The reader is referred
to ref.
\MW\ for all proofs.

\section{The \GD brackets for general $U_1, \ldots U_m$}

In analogy with the scalar case (i.e. $n=1$)
[\ADLER,\LEBMAN,\GELDB,\DIK], I define the second \GD bracket
associated with the $n\times n$-matrix $m^{\rm th}$-order differential
operator $L$ as
$$
\{f,g\}_{(2)}=a\int\tr\ \res \left( L(X_f L)_+ X_g -(L X_f)_+ L X_g
\right)
\eqn\dv$$
where $a$ is an arbitrary scale factor and $X_f, X_g$ are the
pseudo-differential operators
$$\eqalign{
X_f=\sum_{l=1}^m \d^{-l}X_l\quad &, \quad X_g=\sum_{l=1}^m
\d^{-l}Y_l\cr
X_l= \dd f {U_{m+1-l}} \quad &, \quad Y_l= \dd g {U_{m+1-l}} \ .\cr}
\eqn\dvi$$
The functional derivative of $f=\is\tr P(U)$ is defined as usual by
$$\left( \dd f {U_k}(\s) \right)_{ij} =
\sum_{r=0}^\infty \left( -{d\over d\s}\right)^r
\left( {\d \tr P(u)(\s)\over \d (U_k^{(r)})_{ji} }\right)
\eqn\dvii$$
where $(U_k^{(r)})_{ji}$ denotes the $(j,i)$ matrix element of the $r^{\rm
th}$ derivative of $U_k$. It is easily seen, that for $n=1$, equations
\dv-\dvii\ reduce to the standard
 definitions of the \GD brackets [\ADLER,\LEBMAN,\GELDB,\DIK].
 For $m=2,\, n=2$ and
with the extra restrictions $U_1=0,\ \tr\s_3U_2=0$, equation
\dv\  was shown in the previous section to reproduce the
original $V$-algebra \txx\
(with $a=-2\gd$).
Working through the algebra (see [\MW] for details) and defining
 for $l\ge 1$
$$
S_{r,l}^{q,j}=\sum_{s=\max(0,r)}^{\min(q,j)} (-)^{s-r}
\bin s-r+l-1 {l-1} \bin q s \ ,
\eqn\dxiv$$
with
$S_{r,l}^{q,j}=0$ if $\max(0,r)>\min(q,j)$, one obtains the

\noindent
{\bf Proposition 1 :} The second \GD bracket associated with the
$n\times n$-matrix $m^{\rm th}$-order differential operator $L$ as
defined by eqs. \di, \dv\ and \dvi\ equals
$$\eqalign{
\{f,g\}_{(2)}&=a\int\tr \sum_{j=0}^{m-1} \v_j Y_{j+1} \cr
\v_j&=\sum_{l=1}^m \sum_{p=0}^{2m-j-l}
\sum_{q=\max(0,p+j+l-m)}^{\min(m,p+j+l)} \left( S_{q-p,l}^{q,j} -
\bin q-l p \right)
U_{m-q}(X_lU_{m-p-j-l+q})^{(p)}\ .\cr}
\eqn\dxxiii$$

It is not obvious that \dxxiii\ satisfies antisymmetry or the Jacobi identity,
but
this will follow from the Miura transformation discussed below.

\section{The \GD brackets reduced to $U_1=0$}

The problem of consistently restricting a given symplectic manifold
(phase space) to a symplectic submanifold by imposing certain constraints
$\phi_i=0$ has been much studied in the literature. The basic point is
that for a given phase space one cannot set a coordinate to a given
value (or function) without also eliminating the corresponding
momentum. More generally, to impose a constraint $\phi=0$
consistently, one has to make sure that for any functional $f$ the
bracket $\{\phi,f\}$ vanishes if the constraint $\phi=0$ is imposed
{\it after} computing the bracket. In general this results in a
modification of the original Poisson bracket.

Here, I want to impose $\{f, U_1\}\vert_{U_1=0}=0$	for all $f$. Since
$Y_m=\dd g {U_1}$, one sees from \dxxiii\ that this requires
$\v_{m-1}\vert_{U_1=0}=0$. In practice this determines $X_m$ which
otherwise would be undefined if one starts with $U_1=0$. In the
scalar case it is known [\ADLER,\LEBMAN] that $X_m$ should be determined by
$\res [L,X_f]=0$. The following Lemma shows that this is still
true in the matrix case.

\Lemma 2    One has
$\v_{m-1}=-\res [L,X_f]$,
and for
 $U_1=0$, $\res[L,X_f]=0$ is equivalent to
$$X_m={1\over m} \sum_{l=1}^{m-1}\left(
\d^{-1}[U_{m+1-l},X_l]
+\sum_{k=l}^m  (-)^{k-l}\bin k {l-1}
(X_lU_{m-k})^{(k-l)}\right) \ .
\eqn\dxxix$$
Note the commutator term which is a new feature of the present matrix
case as opposed to the scalar case.

One of the main results then is the following

\noindent
{\bf Theorem 3 :}  The second \GD bracket for $n\times n$-matrix
$m^{\rm th}$-order differential operators $L$ with vanishing $U_1$ is
given by
$$\eqalign{
\{f,g\}_{(2)}=&a\int\tr \sum_{j=0}^{m-2} \vt_j Y_{j+1}\ ,  \cr
\vt_j=&{1\over m} \sum_{l=1}^{m-1} [U_{m-j},\d^{-1} [X_l,U_{m-l+1}]]
\cr
+&{1\over m} \sum_{l=1}^{m-1}\Big\{ \sum_{k=0}^{m-l} (-)^k
\bin k+l {l-1} (X_lU_{m-k-l})^{(k)} U_{m-j} \cr
&\phantom{{1\over m} \sum_{l=1}^{m-1}} -\sum_{k=0}^{m-j-1}
\bin k+j+1 j  U_{m-k-j-1} (U_{m-l+1}X_l)^{(k)} \Big\} \cr
+&\sum_{l=1}^{m-1}\,  \sum_{p=0}^{2m-j-l}
\sum_{q=\max(0,p+j+l-m)}^{\min(m,p+j+l)} C_{q-p,l}^{q,j}
U_{m-q}(X_lU_{m-p-j-l+q})^{(p)}\ , \cr
C_{q-p,l}^{q,j}=&S_{q-p,l}^{q,j} -
\bin q-l p -{1\over m}(-)^{q-p+j}\bin q j  \bin p-q+j+l {l-1}\cr }
\eqn\dxxx$$
where the $S_{q-p,l}^{q,j}$ are defined by eq. \dxiv, and it is
understood that $U_0=-1$ and $U_1=0$.

\noindent
{\bf Remark 4 :} If one takes $m=2$, $L=-\d^2+U$, so that
$U_2\equiv U$ and $X_1\equiv X$, only $\vt_0$ is non-vanishing:
$$\vt_0=-{1\over 2} [U,\d^{-1}[U,X]]+{1\over 2}(XU+UX)'
+{1\over 2}(X'U+UX')-{1\over 2}X'''
\eqn\dxxxiii$$
and with $X=\dd f U$ and $Y=\dd g U$ one obtains (using $\int x
\d^{-1} y=-\int (\d^{-1}x) y$)
$$\{f,g\}_{(2)}=a\int\tr \left( -{1\over 2} [U,X]\d^{-1}[U,Y]
+{1\over 2}(X'Y+YX'-XY'-Y'X)U-{1\over 2}YX'''\right)
\eqn\dxxxiv$$
which obviously is a generalization of the original $V$-algebra \txx\
to arbitrary $n\times n$-matrices $U\equiv U_2$.
To appreciate the structure of the non-local terms, I explicitly
write this algebra in the simplest case for $n=2$ (but {\it without}
the restriction
$\tr\s_3U=0$ which is present for \txx). Let
$$U=\pmatrix{ T+V_3&-\rd\vp\cr -\rd\vm& T-V_3\cr} \ .
\eqn\dxxxv$$
Then one obtains from \dxxxiv\ (with $a=-2\gd$) the algebra
$$\eqalign{
\gmd \{T(\s)\, ,\, T(\s')\}  &=
(\d_\s-\d_{\s'})\left[ T(\s') \ds\right]-{1\over 2} \dsppp \cr
\gmd \{T(\s)\, ,\, V^\pm(\s')\}  &=
(\d_\s-\d_{\s'})\left[ V^\pm(\s') \ds\right]\cr
\gmd \{T(\s)\, ,\, V_3(\s')\}  &=
(\d_\s-\d_{\s'})\left[ V_3(\s') \ds\right]\cr
\gmd \{V^\pm(\s)\, ,\, V^\pm(\s')\}&=\es
V^\pm(\s)V^\pm(\s')\cr
\gmd \{V^\pm(\s)\, ,\, V^\mp(\s')\}&=-\es
(V^\pm(\s)V^\mp(\s')+V_3(\s)V_3(\s'))\cr
&\phantom{=}+(\d_\s-\d_{\s'})\left[ T(\s') \ds\right] -{1\over 2}
\dsppp \cr
\gmd \{V_3(\s)\, ,\, V^\pm(\s')\}&=\es
V^\pm(\s)V_3(\s')\cr
\gmd \{V_3(\s)\, ,\, V_3(\s')\}&=\es
(V^+(\s)V^-(\s')+ V^-(\s)V^+(\s'))\cr
&\phantom{=}+(\d_\s-\d_{\s'})\left[ T(\s') \ds\right] -{1\over 2}
\dsppp \ .\cr
}
\eqn\dxxxvi$$
Once again, one sees that $T$ generates the conformal algebra with a
central charge, while $V^+$, $V^-$ and $V_3$ are conformally primary
fields of weight (spin) two. It is easy to check on the example \dxxxiv,
that antisymmetry and the Jacobi identity are satisfied. For general $m$
this follows from the Miura transformation to which I now turn.

\section{The Miura transformation : The case of general $U_1, \ldots U_m$}

\noindent
{\bf Definition and Lemma 5 :}  Introduce the $n\times
n$-matrix-valued functions $P_j(\s),\, j=1, \ldots m$. Then for
functionals $f,g$ (integrals of traces of differential polynomials)
of the $P_j$ the following Poisson bracket is well-defined %
$$\{f,g\}=a\sum_{i=1}^m \int\tr\left( \left( \dd f {P_i} \right)'
\dd g {P_i} - \left[ \dd f {P_i} , \dd g {P_i} \right] P_i \right)
\eqn\ti$$
or equivalently for $n\times n$-matrix-valued (numerical) test-functions
$F$ and $G$
$$\{\int\tr FP_i,\int\tr GP_j\}=a\,  \delta_{ij} \int\tr \left( F'G
-[F,G]P_i \right) \ .
\eqn\tii$$
\noindent
Note that due to the $\delta_{ij}$ in \tii\ one has $m$ decoupled
Poisson brackets. In the scalar case ($n=1$), \tii\ simply gives
$\{P_i(\s),P_j(\sp)\}=(-a)\delta_{ij}\dsp$. These are $m$ free fields
or $m$ $U(1)$ current algebras. In the matrix case, one easily sees that
 each $P_j$ actually gives a $gl(n)$
current algebra. So one has no longer free  fields but $m$ completely
decoupled current algebras. This is still much simpler than the
bracket \dxxiii. To connect both brackets one starts with the
following obvious

\Lemma 6  Let $P_j, j=1, \ldots m$ be as in Lemma 5. Then
$$L=-(\d-P_1)(\d-P_2)\ldots (\d-P_m)
\eqn\tiii$$
is a $m^{\rm th}$-order $n\times n$-matrix linear differential
operator and can be written $L=\sum_{k=0}^mU_{m-k}\d^k$ with $U_0=-1$
as before. This identification gives all $U_k, k=1,\ldots m$ as
$k^{\rm th}$-order differential polynomials in the $P_j$, i.e. it
provides an embedding of the algebra of differential polynomials in
the $U_k$ into the algebra of differential polynomials in the $P_j$.
One has in particular
$$U_1=\sum_{j=1}^m P_j\quad , \quad
U_2=-\sum_{i<j}^m P_iP_j+\sum_{j=2}^m (j-1)P_j'\ .
\eqn\tiv$$

\noindent
I will call the embedding given by \tiii\ a (matrix) Miura
transformation. The most important property of this Miura
transformation is given by the following matrix-generalization of a
well-known theorem [\KW, \DIK].

\noindent
{\bf Theorem 7 :}  Let $f(U)$ and $g(U)$ be functionals of the $U_k,\
k=1,\ldots m$. By Lemma 6 they are also functionals of the $P_j,\
j=1,\ldots m$: $f(U)=\ft(P)$, $g(U)=\gt(P)$ where $\ft(P)=f(U(P))$
etc. One then has
$$\{\ft(P),\gt(P)\}=\{f(U),g(U)\}_{(2)}
\eqn\tv$$
where the bracket on the l.h.s. is the Poisson bracket \ti\ and the
bracket on the r.h.s. is the second \GD bracket \dxxiii.

\noindent
The previous Theorem states that one can either compute
$\{U_k,U_l\}$ using the complicated formula \dxxiii\ or using the
simple Poisson bracket \ti\ for more or less complicated functionals
$U_k(P)$ and $U_l(P)$. In particular Lemma 5 implies the

\noindent
{\bf Corollary 8 :} The second \GD bracket \dxxiii\ obeys
antisymmetry and the Jacobi identity. Bilinearity in $f$ and $g$
being evident, it is a well-defined Poisson bracket.

\section{The Miura transformation : The case $U_1=0$}

As seen from \tiv, $U_1=0$ corresponds to $\sum_{i=1}^m P_i=0$. In
order to describe the reduction to $\sum_i P_i=0$ it is convenient to
go from the $P_i, i=1,\ldots m$ to a new ``basis": $Q=\sum_{i=1}^m
P_i$ and $\P_a, a=1,\ldots m-1$ where all $\P_a$ lie in the
hyperplane $Q=0$. Of course, $Q$ and each $\P_a$ are still $n\times
n$-matrices. More precisely:

\noindent
{\bf Definition and Lemma 9 :}  Consider a $(m-1)$-dimensional
vector space, and choose an overcomplete basis of $m$ vectors $h_j,\
j=1,\ldots m$. Denote the components of each $h_j$ by $h_j^a,\  a=1,
\ldots m-1$. Choose the $h_j$ such that
$$\eqalign{
\sum_{j=1}^m h_j &=0\ ,\cr
h_i\cdot h_j &= \delta_{ij}-{1\over m}\ ,\cr
\sum_{i=1}^m h_i^a h_i^b &=\delta_{ab} \cr}
\eqn\txv$$
and define the completely symmetric rank-3 tensor $D_{abc}$ by
$$D_{abc}=\sum_{i=1}^m h_i^a h_i^b h_i^c \ .
\eqn\txvi$$
Define $Q$ and $\P_a,\  a=1,\ldots m-1$ to be the following linear
combinations of the $P_j$'s
$$\P_a=\sum_{j=1}^m h_j^a P_j \quad , \quad Q=\sum_{j=1}^m P_j\ .
\eqn\txvii$$
If one considers the $P_j$ as an orthogonal basis in a $m$-dimensional
vector-space, then the $\P_a$ are an orthogonal basis in a
$(m-1)$-dimensional hyperplane orthogonal to the line spanned by $Q$.

\noindent
{\bf Proposition 10 :} The Poisson bracket \ti\ can be reduced to the
symplectic submanifold with $Q\equiv \sum_{j=1}^m P_j =0$. The reduced
Poisson bracket is
$$\{f,g\}=a\int\tr \Big(  W_bV_b'
-[V_b,W_c]D_{bcd}\P_d -{1\over m} [V_b,\P_b]\d^{-1}[W_c,\P_c]\Big)
\eqn\txxi$$
where $V_b$ and $W_b$  denote
$V_a= \dd f {\P_a}  , \ V_0= \dd f Q  , \
W_a= \dd g {\P_a}$ and $W_0= \dd g Q $.
Equivalently one has
$$\{\int\tr F\P_b,\int\tr G\P_c\}=a\int\tr \Big(  GF'\delta_{bc}
-[F,G]D_{bcd}\P_d -{1\over m} [F,\P_b]\d^{-1}[G,\P_c]\Big) \ .
\eqn\txxii$$
\noindent
{\bf Corollary 11 :}  The Poisson bracket \tii\ when reduced to
$Q=\sum_{j=1}^mP_j=0$ can be equivalently written as
$$\eqalign{
\{f,g\}=a\int\tr &\Big(  GF'(\delta_{ij} -{1\over m}) -
[F,G] (\delta_{ij} -{2\over m}){1\over 2}(P_i+P_j)\cr
&-{1\over m} [F,P_i]\d^{-1}[G,P_j]\Big) \ . \cr}
\eqn\txxv$$

\noindent
{\bf Theorem 12 :}   Let $U_1=0$ and hence $Q=\sum_{j=1}^mP_j=0$.
By the Miura transformation of Lemma 6 any functionals $f(U),\, g(U)$
of the $U_k$ only $(k=2,\ldots m$)  are also functionals
$\ft(\P)=f(U(\P)),\, \gt(\P)=g(U(\P))$ of the $\P_a, a=1, \ldots m$
only. The reduced second \GD bracket \dxxx\ of $f$ and $g$ equals the
reduced Poisson bracket \txxi\ of $\ft$ and $\gt$.

\noindent
{\bf Corollary 13 :}  The second \GD bracket \dxxx\ obeys
antisymmetry and the Jacobi identity. Bilinearity in $f$ and $g$ being
evident, it is a well-defined Poisson bracket.

\section{The conformal properties}

\REF\MAT{P. Mathieu, {\it Extended classical conformal algebras and
the second Hamiltonian structure of Lax equations}, Phys. Lett. {\bf
B208} (1988) 101.}
\REF\DIZ{P. Di Francesco, C. Itzykson and J.-B.
Zuber, {\it Classical $W$-algebras}, Commun. Math. Phys.
{\bf 140} (1991) 543 .}
In the scalar case, i.e. for $n=1$, the second \GD bracket (with
$U_1=0$) gives the $W_m$-algebras [\BAK,\MAT,\DIZ]. The interest in the
$W$-algebras stems from the fact that they are extensions of the
conformal Virasoro algebra, i.e. they contain the Virasoro algebra as a
subalgebra. Furthermore, in the scalar case, it is known that certain
combinations of the $U_k$ and their derivatives yield primary fields of
integer spins $3,4,\ldots m$. It is the purpose of this section to
establish the same results for the matrix case, $n>1$. From now on,
I only consider the second \GD bracket \dxxx\ for the case
$U_1=0$. I will simply write $\{f,g\}$ instead of $\{f,g\}_{(2)}$.
Also, it is often more convenient to replace the scale factor $a$ by
$\gd$ related to $a$ by
$$a=-2\gd\ .
\eqn\qi$$
(Note that $\gd$ need not be positive.)

\subsection{The Virasoro subalgebra}

For the original $V$-algebra \txx\  (corresponding to
$m=2,\, n=2$ and an additional constraint $\tr\s_3U_2=0$) one sees that
$T={1\over 2}\tr U_2$ generates the conformal algebra. I will now show
that for general $m, n$, the generator of the conformal algebra is
still given by this formula.

\Lemma 14   For arbitrary $m\ge 2$ one has
$$\eqalign{
\{\int\tr FU_2,\int\tr GU_2\}&
= a\int\tr \Big(
-{1\over m} [F,U_2]\d^{-1}[G,U_2]-[F,G](U_3-{m-2\over 2}U_2')\cr
&+{1\over 2}(F'G-G'F+GF'-FG')U_2 -{1\over 2} \bin m+1 3  G F''' \Big) \
. \cr}
\eqn\qii$$
Note that for $m=2$ one has to set $U_3=0$.

\noindent
{\bf Proposition 15 :}  Let $T(\s)={1\over 2} \tr U_2(\s)$. Then
$$\gmd \{T(\s_1),T(\s_2)\}=(\d_{\s_1}-\d_{\s_2})
\left( T(\s_2)\delta(\s_1-\s_2)\right)
-{n\over 4}\bin m+1 3 \delta'''(\s_1-\s_2) \ .
\eqn\qiv$$
Equivalently, if, for $\s\in S^1$, one defines for integer $r$
$$L_r=\gmd\int_{-\pi}^\pi \rmd\s\,  T(\s)e^{ir\s} +{c\over
24}\, \delta_{r,0}
\eqn\qv$$
where
$$c={6\pi\over \gd} \, n\, \bin m+1 3 = {12\pi\over (-a)} n\, \bin m+1 3
\eqn\qvi$$
then the $L_r$ form a Poisson bracket version of the Virasoro algebra
with (classical) central charge $c$ :
$$i\{L_r,L_s\}=(r-s)L_{r+s}+{c\over 12} (r^3-r) \delta_{r+s,0} \ .
\eqn\qvii$$
Also, if $\{A_\mu\}_{\mu=1,\ldots n^2-1}$ is a basis for the traceless
$n\times n$-matrices, then each $S_\mu (\s)=\tr A_\mu U_2(\s),
\mu=1,\ldots n^2-1$ is a conformally primary field of conformal
dimension (spin) 2: %
$$\gmd \{T(\s_1),S_\mu(\s_2)\}=(\d_{\s_1}-\d_{\s_2})
\left( S_\mu(\s_2)\delta(\s_1-\s_2)\right)
\eqn\qviii$$
or for the modes $(S_\mu)_r=\gmd\int_{-\pi}^\pi \rmd\s\,
S_\mu(\s)e^{ir\s}$ one has
$i\{L_r,(S_\mu)_s\}=(r-s)(S_\mu)_{r+s}$. Equations \qiv\ and \qviii\
can be written in matrix notation as ($\id$ denotes the $n\times n$
unit matrix)
$$\gmd \{T(\s_1),U_2(\s_2)\}=(\d_{\s_1}-\d_{\s_2})
\left( U_2(\s_2)\delta(\s_1-\s_2)\right)
-{1\over 2}\bin m+1 3 \id\,  \delta'''(\s_1-\s_2) \ .
\eqn\qix$$

\subsection{The conformal properties of the $U_k$ for $k\ge 3$}

Having computed the conformal properties of
the matrix elements of $U_2$, I will now give
those of all the other $U_k$, i.e. compute $\{T(\s_1),U_k(\s_2)\}$ or
equivalently, for any (test-) function $\e(\s)$, compute $\{\int \e T,
U_k(\s_2)\}$
for all $k\ge 3$. I will find that this Poisson bracket is linear in the
$U_l$ and their derivatives and is formally identical to the result of
the scalar case. It then follows that appropriately symmetrized
combinations $W_k$ can be formed that are $n\times n$-matrices, each
matrix element of $W_k$ being a conformal primary field of dimension
(spin) $k$.

\noindent
{\bf Proposition 16 :}
The conformal properties of all matrix elements of all $U_k, k=2,\ldots
m$ are given by
$$\eqalign{
\gmd \{\int \e T, U_k\} &= -\e U_k'-k\e' U_k +{k-1\over 2} \bin m+1 {k+1}
\e^{(k+1)}\cr
& +\sum_{l=2}^{k-1}\left[ \bin m-l {k+1-l} -{m-1\over 2} \bin m-l {k-l}
\right] \e^{(k-l+1)} U_l \cr }
\eqn\qxv$$
which is formally the same equation as in the scalar case $n=1$ [\DIZ].

Since the conformal properties \qxv\ are formally the same as in the
scalar case, and in the latter case it was possible to form
combinations $W_k$ that are spin-$k$ conformally primary fields [\DIZ], one
expects a similar result to hold in the matrix case. Indeed, one has the

\noindent
{\bf Theorem 17 :}   For matrices $A_1, A_2, \ldots A_r$ denote by
$S[A_1,A_2,\ldots,A_r]$ the completely symmetrized product normalized
to equal $A^r$ if $A_s=A$ for all $s=1,\ldots r$. Let
$$\eqalign{
W_k=&\sum_{l=2}^k B_{kl\, }U_l^{(k-l)} +\sum_{0\le p_1\le \ldots \le p_r
\atop \sum p_i+2r=k} (-)^{r-1} C_{p_1\ldots p_r}\,  S[U_2^{(p_1)},
\ldots, U_2^{(p_r)}]\cr
&+\sum_{0\le p_1\le \ldots \le p_r
\atop s\le l\le k-\sum p_i-2r} (-)^{r} D_{p_1\ldots p_r,l}\,
S[U_2^{(p_1)}, \ldots, U_2^{(p_r)},U_l^{(k-l-\sum p_i-2r)}]\cr }
\eqn\qxix$$
where the coefficiets $B_{kl}, C_{p_1\ldots p_r}$ and $D_{p_1\ldots
p_r,l}$ are the same as those given in ref. \DIZ\ for the scalar case, in
particular
$$B_{kl}=(-) ^{k-l}\ { \bin k-1 {k-l} \bin m-l {k-l} \over
\bin 2k-2 {k-l} }\ .
\eqn\qxx$$
Then the $W_k$ are spin-$k$ conformally primary
$n\times n$-matrix-valued fields, i.e.
$$\gmd \{\int \e T, W_k\}=-\e W_k' -k\e' W_k \ .
\eqn\qxxi$$
For $\s\in S^1$ one can define the modes
$(W_k)_s=\gmd\int_{-\pi}^\pi \rmd\s\, W_k(\s) e^{is\s}$ and the Virasoro
generators $L_r$ as  in \qv. Then one has equivalently
$$i\{L_r,(W_k)_s\}=\big( (k-1)r-s\big) (W_k)_{r+s}
\eqn\qxxia$$
where each $(W_k)_s$ is a $n\times n$-matrix.

\noindent
{\bf Examples :}   From the previous Theorem and the results of ref.
\DIZ\ (their Table I) one has explicitly:
$$\eqalign{
W_3&= U_3-{m-2\over 2} U_2' \cr
W_4&= U_4-{m-3\over 2} U_3'+{(m-2)(m-3)\over 10} U_2''
+{(5m+7)(m-2)(m-3)\over 10 m (m^2-1)} U_2^2 \cr
W_5&= U_5-{m-4\over 2} U_4'+{3(m-3)(m-4)\over 28} U_3''
-{(m-2)(m-3)(m-4)\over 84} U_2'''\cr
&+{(7m+13)(m-3)(m-4)\over 14 m (m^2-1)} (U_2W_3+W_3U_2) \ . \cr}
\eqn\qxxiii$$

\section{Example of the $V_{n,3}$-algebra}

{}From the previous subsection one might have gotten the impression that
the matrix case is not very different from the scalar case. This is
however not true. In the previous subsection only the conformal
properties, i.e. the Poisson brackets with $T={1\over 2} \tr \id U_2$
were studied, and since the unit-matrix $\id$ always commutes, most of
the new features due to the non-commutativity of matrices were not
seen. Technically speaking, only $\tr \vt$ was needed, not $\vt$
itself (cf. eq. \dxxx). In this subsection, I will give the Poisson brackets,
for the (more
interesting) reduction to $U_1=0$, of any two matrix elements of $U_2$
or $U_3$, or equivalently of $U_2$ or $W_3$, for  $m=3$.
 This is the complete algebra, giving a matrix generalization
\REF\ZAM{A.B. Zamolodchikov, {\it Infinite additional symmetries
in two-dimensional conformal field theory}, Theor. Math. Phys. {\bf 65}
(1985) 1205. }
of Zamolodchikov's $W_3$-algebra [\ZAM]. (Recall that $F$ and $G$ are $n\times
n$-matrices of
test-functions.)
$$\eqalign{
\{\int\tr FU_2,\int\tr GU_2\}=a \int\tr\Big(
&-{1\over 3}[F,U_2]\d^{-1}[G,U_2] -[F,G]W_3\cr
&+{1\over 2}(F'G+GF'-FG'-G'F)U_2-2GF'''\Big)\ ,
\cr}
\eqn\uiii$$
$$\eqalign{
\{\int\tr FU_2,\int\tr GW_3\}=a \int\tr \Big(
&-{1\over 3}[F,U_2]\d^{-1}[G,W_3] -{1\over 6}[F,G]U_2^2\cr
&+(-{1\over 4}[F',G']+{1\over 2}[F'',G]+{1\over 12}[F,G'']) U_2\cr
&+(F'G+GF'-{1\over 2}FG'-{1\over 2}G'F)W_3 \Big)\ ,
\cr}
\eqn\uiv$$
$$\eqalign{
\{\int\tr F&W_3,\int\tr GW_3\}
=a \int\tr \Big(
-{1\over 3}[F,W_3]\d^{-1}[G,W_3]\cr
&-{1\over 6}[F,G](W_3U_2+U_2W_3)
+{2\over 3}(FU_2GW_3-GU_2FW_3)\cr
&+{5\over 12}(F'U_2GU_2-G'U_2FU_2)+{1\over 12}(FG'-GF')U_2^2
+{1\over 12}[F,G]U_2'U_2\cr
&+{7\over 12}[F',G']W_3-{1\over 6}[F,G]''W_3
+{1\over 12}(FG'''+G'''F-F'''G-GF''')U_2\cr
&+{1\over 8}(F''G'+G'F''-F'G''-G''F')U_2+{1\over 6}GF^{(5)} \Big)\ .
\cr}
\eqn\uv$$

One remarks that in the scalar case ($n=1$) this reduces to the
Poisson bracket version of Zamolodchikov's $W_3$-algebra [\ZAM], as it
should. In the matrix case however, even if $F=f\id,\ G=g\id$ (with
scalar $f,g$) this is a different algebra, i.e.
$\{\int\tr W_3(\s),\int\tr W_3(\sp)\}$ does not reduce to the
$W_3$-algebra, since the r.h.s. contains the non-linear terms and
$\tr U_2^2\ne (\tr U_2)^2$. In other words, the scalar ($n=1$)
$W_m$-algebras are not subalgebras of the matrix $V_{n,m}$-algebras.
The only exception is $m=2$, since one always has a Virasoro subalgebra.

\refout

\end